\documentclass[twocolumn,prl,showpacs]{revtex4}

\usepackage{graphicx}
\usepackage{rotating}
\usepackage{amsmath}
\usepackage{amsfonts}
\usepackage{amssymb}
\usepackage{enumerate}
\usepackage{longtable}
\usepackage{dcolumn}
\usepackage{bm}

\begin{document}

\newcommand{\be}{\begin{equation}}
\newcommand{\ee}{\end{equation}}
\newcommand{\bn}{\begin{eqnarray}}
\newcommand{\en}{\end{eqnarray}}

\title{Valence Fluctuation Driven Non-FL behavior and Unconventional   
Superconductivity in Pressurised $CeCu_{2}Si_{2}$.}

\author{M. S. Laad}

\affiliation{Max-Planck Institut f\"ur Physik Komplexer Systeme, 
N\"othnitzer Strasse 38, D01187 Dresden, Germany}

\date{\today}

\begin{abstract}
   I study an Extended Periodic Anderson Model (EPAM)
with non-local hybridisation, $V_{fc}$, and a coulomb interaction, $U_{fc}$,
 between localised $f$ electrons and wide
band conduction ($c$) electrons.   
Within DMFT,
a quantum phase transition (QPT) driven by soft $f$-valence fluctuations,
accompanied by a discontinuous jump
in the Fermi volume, is found as $V_{fc}$ is tuned across a critical value. 
Near the associated QCP, an unconventional superconductive instability,
driven by these extended, singular valence fluctuations, naturally emerges.  
{\it All} these 
phenomena are representative of observations in pressurised $CeCu_{2}Si_{2}$.
\end{abstract}


\maketitle


  Rare-earth (RE) intermetallics continue to throw up ever new surprises.
Evidences of non-Fermi liquid (nFL) behavior near magnetic transitions~\cite{[1],[2]}, frequently
accompanied by (unconventional) superconductivity (SC) present challenges to 
theory.  Given that the ``standard model'' of $f$-band systems, the Periodic 
Anderson Model (PAM), emphasizes the heavy FL phenomenon driven by {\it quasilocal} Kondo screening~\cite{[3]}, these new observations call for  
mechanisms which destabilize Kondo singlet formation.  While a lot of  
work focuses on the role of critical magnetic fluctuations- local or
 extended~\cite{[2],[4],[5],[6],[7]}, in this context, much less effort 
has been expended in investigating 
alternative mechanisms involving charge degrees of freedom.  This 
is partly due to the paucity of new materials 
which require introduction of such additional, exotic scenarios.  

  Among the most striking examples of materials where such an eventuality is 
necessary (Fig.(1) of Ref.[8]) is the 
cerium ($Ce$)-based HF-SC $CeCu_{2}Si_{2}$~\cite{[8]}.  
As a function 
of external pressure ($p$), it shows {\bf two} QPTs: at $p_{c1}$, incommensurate
 SDW order is continuously destroyed, and SC setting in at lower $T$ masks the $T=0$ AF-QCP.  While this nFL metal is characterisable by the spin-fluctuation based theories for antiferromagnetic (AFM) QCPs, the
second QPT at $p_{c2}$ presents an enigma.  Explicitly, observation of 
a resistivity, $\rho_{dc}(T) \simeq T$ and its invariance with disorder 
is inexplicable within the $D=3$ SDW-QCP
scenario, but seemingly in agreement with {\it local}
valence fluctuations.  Surprisingly, {\it enhanced} SC emerges at 
$p>p_{c1}$, remaining stable over a wide pressure range ($\Delta p>2.0$ GPa), and peaking around $p_{c2}$, before being cut off rapidly beyond.  Concomitantly,
nFL behavior for $p<p_{c2}$ goes over to a FL for low $T$.  These observations 
pose a challenge for the standard model, and 
have been largely unaddressed,
rendering an understanding of the two QPTs in $CeCu_{2}Si_{2}$ an
open, unsolved problem.  

 Existence of the QPT at $p_{c2}$ also raises new questions about the nature of
the AF-QCP at $p_{c1}$, since the nFL phase extends {\it continuously} from 
$0<p<p_{c2}$.  What role does the QPT at $p_{c2}$ play in the emergence of 
SDW-AF at $p<p_{c1}$?  Is there a crossover from a {\it local} QPT at $p_{c2}$
to a $z=2$ QC region for $p_{c1}<p<p_{c2}$, and, if so, how might this come about?  Shedding light on these issues 
mandates a detailed study of the nFL state for $p\simeq p_{c2}$.  Here, 
soft valence fluctuations (VF) are relevant, as evidenced by a $Ce$-like volume collapse across $p_{c2}$.  A valence
 fluctuation (VF) mechanism has indeed been speculated to drive the nFL behavior and SC
around 
$p_{c2}$~\cite{[9]}.  
It may also have broader relevance to other $Ce$-based
compounds~\cite{[10]}.

  On general theoretical grounds, this raises several pertinent issues which a 
theoretical model must address:  

(i)  In contrast to the usual PAM, little is known in detail about how Kondo
``quenching'' occurs with extended hybridisation.  What is the role of the 
nodes of $V_{fc}({\bf k})$ in this context?  

(ii)  Can extended hybridisation destabilise a heavy FL state?  
What 
is the detailed nature of the resulting QCP/QPT? 
Does the Fermi volume jump 
abruptly across the QPT?   

(iii)  Given nodes in $V_{fc}({\bf k})$, any superconductivity should be of the 
unconventional type.  How might this occur?   Are additional, unconventional 
(e.g, density wave) states also possible?  If so, what is their relation to 
SC?

 Here, I address these issues by proposing a modified  
 PAM with extended $f$-hopping and hybridisation,
 as well as a direct coulomb interaction between 
the $f$ electrons and conduction ($c$) electrons, dubbed Extended-PAM (EPAM).  
 The Hamiltonian is $H=H_{0}+H_{1}$, with the itinerant part described by

\be
H_{0}=-t_{f}\sum_{<i,j>,\sigma}f_{i\sigma}^{\dag}f_{j\sigma} - t_{p}\sum_{<i,j>,\sigma}c_{i\sigma}^{\dag}c_{j\sigma} + V_{fc}\sum_{<i,j>,\sigma}f_{i\sigma}^{\dag}c_{j\sigma} 
\ee
and the local part, by

\be
H_{1}= U_{ff}\sum_{i}n_{if\uparrow}n_{if\downarrow} + U_{fc}\sum_{i,\sigma,\sigma'}n_{if\sigma}n_{ic\sigma'}  
+ \epsilon_{f}\sum_{i}n_{fi}
\ee 
I take the $c$-band centered around $E=0$ and consider
$U_{ff}=\infty$ (valid for $f$ shells), so the $f$ electrons are 
projected fermions, $X_{if\sigma}=(1-n_{if-\sigma})f_{i\sigma}$,
satisfying $[X_{i\sigma},X_{j\sigma'}^{\dag}]_{+}=\delta_{ij}\delta_{\sigma\sigma'}(1-n_{if,-\sigma})$. 
Using the Gutzwiller approximation, one replaces $X_{if\sigma}=q_{\sigma}f_{i\sigma}$ with $q_{\sigma}=(1-n_{f})/(1-n_{f\sigma})$ and $n_{f\sigma}=(1/N)\sum_{i}\langle f_{i\sigma}^{\dag}f_{i\sigma}\rangle$.
This implies $(t_{f},\epsilon_{f})\rightarrow q_{\sigma}^{2}(t_{f},\epsilon_{f})$ and 
$V_{fc}\rightarrow q_{\sigma}V_{fc}$ in what follows.  This corresponds to the
slave-boson mean-field theory (SB-MFT), and yields a narrow, coherent $f$ band
with a width $W\simeq k_{B}T_{K}^{mf}$~\cite{[7]}, the mean-field Kondo scale.
 The effect of external pressure is simulated by 
varying $\epsilon_{f}$ and $V_{fc}$ in $H$ above, and I take $\epsilon_{f}\simeq\mu$ in what follows.  In particular, I investigate the fate of this SB-MFT 
Kondo scale in presence of strong, quantum fluctuations of the $f$ valence.
  
  I start by splitting the $V_{fc}$ term as
$(V_{fc}-\sqrt{t_{f}t_{p}})\sum_{<i,j>,\sigma}(f_{i\sigma}^{\dag}c_{j\sigma}+h.c)
+\sqrt{t_{f}t_{p}}\sum_{<i,j>,\sigma}(f_{i\sigma}^{\dag}c_{j\sigma}+h.c)$
and consider $H=H_{0}+H_{1}$ with $V_{fc}=\sqrt{t_{f}t_{p}}$ to begin with.
Employing the combinations,
$a_{i\sigma}=(uX_{if\sigma}+vc_{i\sigma}), b_{i\sigma}=(vX_{if\sigma}-uc_{i\sigma})$
with $u=\sqrt{t_{f}/(t_{f}+t_{p})}, v=\sqrt{t_{p}/(t_{f}+t_{p})}$.
it is easy to see that $H=H_{0}+H_{1}$ is
$H_{0}=-t\sum_{<i,j>,\sigma}(a_{i\sigma}^{\dag}a_{j\sigma}+h.c)$ and
$H_{1}= U_{fc}\sum_{i,\sigma,\sigma'}n_{ia\sigma}n_{ib\sigma'} + \epsilon_{f}\sum_{i,\sigma}[n_{ia\sigma}+n_{ib\sigma}+(a_{i\sigma}^{\dag}b_{i\sigma}+h.c)]$.

  This is the spin $S=1/2$ Falicov-Kimball model (FKM) with a 
{\it local} hybridisation term, which is finite whenever $\epsilon_{f}\ne 0$.   
Remarkably, when $\epsilon_{f}=0$, this 
reduces to the pure $S=1/2$ FKM!  Below, we show how $(\epsilon_{f}=0,V_{fc}=\sqrt{t_{f}t_{p}})$ 
separates {\it two}, different metallic phases.  

  First, at $\epsilon_{f}=0$, we see that $[n_{ib},H]=0$ for {\it each} $i$,
implying a {\it local} $U(1)$ invariance of $H$: local configurations with
$n_{b}=0,1$ are rigorously degenerate.  This is {\it exactly} the condition for
soft VF to emerge.  As is known~\cite{[11]}, the symmetry
unbroken metallic phase is consequently {\it not} a FL, but is dominated by a 
superposition of one-particle ($n_{b}=0$) and two-particle ($n_{b}=1$) states
at low energy.  The $a$-fermion spectral function 
is, assuming a lorentzian unperturbed DOS for analytical clarity, very simple:

\be
\rho_{a}(\omega)=\frac{1-n_{b}}{\omega+iW} + \frac{n_{b}}{\omega-U_{fc}+iW}
\ee
where $W$ is the half-width of the lorentzian, $\rho_{0}(\epsilon)=(W/\pi)(\epsilon^{2}+W^{2})^{-1}$.  The corresponding $b$-fermion propagator has branch cut
singular behavior precisely at $\mu(=0)$, leading to singularities in the 
local,  one- and two-particle responses: $\rho_{b}(\omega) \simeq \theta(\omega)|\omega|^{-(1-\alpha)}$ and $\chi_{ab}"(\omega)=\int dt e^{i\omega t}\langle a_{i\sigma}^{\dag}b_{i\sigma}(t);b_{i\sigma}^{\dag}a_{i\sigma}(0)\rangle \simeq \theta(\omega)|\omega|^{-(2\alpha-\alpha^{2})}$.  Here, $\alpha=(1/\pi)$tan$^{-1}(U_{fc}/W)$ is the so-called $s$-wave phase shift of the Anderson-Nozieres-de Dominicis (AND) X-Ray Edge (XRE) problem.  Obviously, the FL quasiparticle 
weight, $Z=0$, and the $\omega, T$ dependence of physical quantities will
be governed by power-law responses.  Notice that, in the $(f,c)$ basis,
 these divergences correspond to extended (cf. non-local hybridisation),
singular quantum fluctuations of the $f$
valence: it is precisely these fluctuations which
destroy FLT at $\epsilon_{f}=0$ via the 
AND orthogonality catastrophe (OC)~\cite{[12]}.

  In our FKM with $\epsilon_{f}=0$, and at $T=0$, 
the total fermion number, $n=n_{a}+n_{b}=n_{f}+n_{c}$, jumps discontinuously 
from $n_{-}=(1/2)+(1/\pi)tan^{-1}(U_{fc}/2W)$ to $n_{+}=(3/2)-(1/\pi)tan^{-1}(U_{fc}/2W)$ for a range of densities, $n$, near unity.  This corresponds to a 
sudden jump in the $f$-occupation, $n_{f}$, 
giving a {\it first} order 
valence transition as $\epsilon_{f}$ is tuned through $\mu(=0)$. 
At finite $T$, this line of first-order transitions ends at a 
second order quantum critical end-point (QCEP) as in Fig.(1), and
$n_{f}$ varies very rapidly over an energy scale $O(k_{B}T)$ 
around $\epsilon_{f}=0$, 
extrapolating to a jump $T=0$.  Notice that this jump in $n_{f}$ depends on 
$U_{fc}/t$, which we choose henceforth to be such that the system is close to 
this QCEP.
Simultaneously,
the compressibility, $\chi=dn/d\mu$, remains
finite, and thus no electronic phase separation results.  Within DMFT, this 
explicitly shows the link between emergence of singular non-FL behavior and
the (selective Mott) 
localisation-delocalisation transition of the $f$-electrons.  In contrast 
to the usual PAM with {\it local} hybridisation, we 
have shown this link explicitly in the EPAM at a {\it finite} $V_{fc}$.  Recent slave boson approaches 
have proposed this ``selective Mott transition'' of $f$-electrons in the context of the QCPs in RE systems~\cite{[7]}.  Our work is a concrete, DMFT-based,
realization of the ``selective Mott'' QPT, 
and here, nFL behavior arises from the
AND-OC in the corresponding impurity problem as the extended hybridisation is
varied across a critical value, or as the system is tuned across a critical 
pressure, $p_{c}$.   
  
  To proceed, consider the bosonised lagrangian of the XRE model
corresponding to $H$ in the local limit of the lattice model.  
Following Schotte {\it et al.}~\cite{[13]}, this is one-dimensional in {\it each} radial direction around the local site, and reads $L_{0}=L_{0}'+L_{0}"$, 
where

\be
L_{0}'=\frac{u_{\rho,\sigma}}{2}\sum_{\rho,\sigma}\int [K_{\rho,\sigma}\Pi_{\rho,\sigma}^{2}(r)+\frac{1}{K_{\rho,\sigma}}(\partial_{r}\phi_{\rho,\sigma}(r))^{2}]dr
\ee
and
$L_{0}"=\frac{g}{\pi u}\sum_{\rho}\int \partial_{r}\phi_{\rho}(r)dr$.  Here,
$u_{\rho,\sigma},K_{\rho,\sigma}$ are 
explicit functions of $U_{fc}/t$.  
 Introducing the usual symmetric-antisymmetric (charge-spin) combinations of
$\phi_{\rho,\sigma}(r)$, we see that the antisymmetric (spin) channel
 completely decouples from the charge channel: 
a kind of high-dimensional 
{\bf spin-charge separation}! 
Its underlying origin is the AND OC.  Thus, the physical response
will be completely governed by the critical (power-laws at low energy)
spin and charge collective modes.  
  Further, $\sum_{i,\sigma,\sigma'}U_{fc}^{\sigma\sigma'}n_{if\sigma}n_{ic\sigma'}$ with 
$U_{fc}^{\sigma,-\sigma}<U_{fc}^{\sigma,\sigma}$  
implies that $\chi_{\sigma,-\sigma}''(\omega) > \chi_{\sigma,\sigma}''(\omega)$, i.e, 
that the dominant superconductive instabilities driven by singular VF
at lower $T$ 
will be in the 
spin-singlet sector.  

  Spin-charge separation sets both free to order independently.  
One would then observe {\it two} QPTs, associated with spin (AF)
ordering and valence fluctuations (charge).  Given the ${\bf k}$-dependence of
$V_{fc}$, e.g, if $V_{fc}({\bf k})\simeq$ (cos$k_{x}$-cos$k_{y}$) in $D=2$, the
charge sector could develop $d$-wave instabilities in the nFL phase.

  Consider now $\epsilon_{f}\ne 0$.  This has two effects: (i) it moves the 
$b$-fermion level away from $\omega=0$, and, 
(ii) the $a$-$b$ hybridisation implies
generating a finite, but heavy $b$-fermion mass, due to recoil in the XRE 
problem~\cite{[15]}.  This leads to a small ``coherence scale'', the recoil energy, $\epsilon_{rec}=k_{B}T_{coh}$,
 below which HFL metallic behavior obtains in the corresponding 
lattice model.  The FL quasiparticle overlap, $Z \simeq e^{-C(t=\infty)}$,   
with $C(t)=2U_{fc}^{2}\int \frac{\chi_{ab}"(\omega)}{\omega^{2}}(1-cos(\omega t))d\omega$.  This gives $Z \simeq exp[U_{fc}^{2}(ln(\gamma)/(1-\gamma^{2}))]$.  Here, $\gamma=m_{a}/m_{b}$, with $m_{a}$ the band mass of
the $a$-fermion and $m_{b}$ the heavy mass of the $b$ fermion. 
Hence $T_{coh} \propto Z$ (note the difference from the SBMFT 
scale, $T_{K}^{mf}$, due to strong VF)
increases with $\epsilon_{f}$, as indeed observed in the
region to the right of the VF-QPT.

  This is a specific, soluble example of a non-FL: at $\epsilon_{f}=0$, the strict vanishing of the hybridisation ($\epsilon_{f}$) 

(i) destroys the Kondo effect, as reflected in the absence of the narrow FL
resonance in the DOS of the $D=\infty$ FKM, and

(ii) gives power-law
 fall-off in the local one- and two-particle responses describing singular
hybridisation fluctuations ($\chi_{\sigma,\sigma'}(\omega)$).

 and a HFL when $\epsilon_{f}\ne 0$.  Interestingly, this  
arises via competition between $V_{fc}$ and $U_{fc}$, and involves collective
{\it density} fluctuations, rather than the usual spin fluctuations: 
they play a {\it dual} role, causing both, the heavy FL
and destroying it across a critical $V_{fc}=\sqrt{t_{f}t_{p}},\epsilon_{f}=0$. 
This is very different from the canonical picture, and constitutes a 
new route to HFL behavior.

  This loss of
one-particle coherence at $\epsilon_{f}=0$ facilitates instabilities toward
(two-particle) ordered states.   
As in coupled $1D$ Luttinger liquids~\cite{[16]}, 
this high-$D$ nFL is now unstable, either toward
 unconventional SC/density-wave order, or toward a correlated FL metal, away
 from 
this {\it unstable} fixed point.
 
  Consider first $\delta V_{fc}=(V_{fc}-\sqrt {t_{f}t_{p}})>0$.  The extra
term, $\sum_{<i,j>,\sigma}\delta V_{fc}(f_{i\sigma}^{\dag}c_{j\sigma}+h.c)$ also causes one-particle {\it intersite} hybridisation between the 
$a,b$ fermions, producing extra terms like
$H_{ab}^{(0)}=\delta V_{fc}\sum_{<i,j>,\sigma}[\alpha_{fp}(a_{i\sigma}^{\dag}a_{j\sigma}-b_{i\sigma}^{\dag}b_{j\sigma}) + \beta_{fp}(a_{i\sigma}^{\dag}b_{j\sigma}+h.c)]$,
with $\alpha_{fp}=\frac{\sqrt{t_{f}t_{p}}}{t_{f}+t_{p}}$, and 
$\beta_{fp}=\frac{t_{p}-t_{f}}{t_{f}+t_{p}}$.
This gives the $b$-fermions a finite mass and suppresses the 
infra-red singularities found for $V_{fc}=\sqrt {t_{f}t_{p}}, \epsilon_{f}=0$,
resulting in another HFL with $T_{coh} \propto Z<<1$.  For $\delta V_{fc}<0$, low-energy p-h bound states ($\langle a_{i\sigma}^{\dag}b_{j\sigma}\rangle>0$) drive
 an unconventional (nFL) density-wave (UDW) metal. The one-electron dispersion 
is modified to 
$E_{\bf k}=\pm\sqrt{\epsilon_{\bf k}^{2}+\delta V_{fc}^{2}({\bf k})}$, leading to a low-energy pseudogap in the DMFT propagators, in much the same way as in the $D=\infty$ Hubbard
 model in the AF metal phase~\cite{[17]}.  In the EPAM, this PG is associated with 
the UDW state, as shown below.

Given the singular VF propagator, $\chi_{ab}"(\omega)$, found above,
an instability to unconventional superconductivity
can occur via exchange of these soft collective VF modes, as
studied in earlier~\cite{[9],[gil]} work. 
  Here, I use high-$D$ bosonisation to study the instabilities of this 
non-FL state to
UDW/SC order.

  For small $\epsilon_{f}, \delta V_{fc}$, 
when the one electron $a-b$ ``hybridisation'' 
triggers {\it coherent} two-particle hopping processes either in the 
particle-hole (ph) or the particle-particle (pp) channels via terms like
$H^{(2)}=\sum_{<i,j>,\sigma,\sigma'}g_{\sigma\sigma'}a_{i\sigma}^{\dag}b_{j\sigma}b_{j\sigma'}^{\dag}a_{i\sigma'}$, the {\it impurity} lagrangian (obtained after decoupling $H^{(2)}$ in $ph$ and $pp$ channels) in 
the charge sector in the presence of spin-charge separation is
$L_{\rho}=L_{0,\rho}+L_{1,\rho}$ with $L_{0,\rho}$ from Eq.(4) and
$L_{1,\rho}= \int [g_{1}cos(\sqrt{8\pi K_{\rho}})\phi_{\rho} + g_{2}cos(\sqrt{8\pi/K_{\rho}})\theta_{\rho}]dr$.
 
The spin sector is described by another sine-Gordon model, $L_{\sigma}=L_{0,\sigma}+L_{1,\sigma}$ with $L_{0,\sigma}$ from Eq.(4) and
$L_{1,\sigma}=- g_{\sigma}\int cos(\sqrt{8\pi K_{\sigma}})\phi_{\sigma}]dr$.
 Here, $g_{\alpha},\alpha=1,2,\rho,\sigma$ and $K_{\rho,\sigma}$ are functions 
of $U_{fc},V_{fc},\epsilon_{f},W$.  Assuming a disordered {\it paramagnetic}
state, the $\sigma$-QsG model is off-critical.  
Focussing on $L_{\rho}=L_{0,\rho}+L_{1,\rho}$, 
a conventional RG procedure describes the renormalization flow of the charge ($\rho$) sector:

\be
dg_{1}/dl=2(1-K_{\rho})g_{1}+ \epsilon_{f}^{2}(K_{\rho}-K_{\rho}^{-1})
\ee
with

\be
dg_{2}/dl=2(1-K_{\rho}^{-1})g_{2} - \epsilon_{f}^{2}(K_{\rho}-K_{\rho}^{-1})
\ee
and

\be
d (logK_{\rho})/dl=\frac{1}{2}(K_{\rho}^{-1}g_{2}^{2}-K_{\rho}g_{1}^{2})
\ee

similar to those derived earlier in 
other contexts~\cite{[18]}.  Analysis of these equations 
shows that, depending upon whether $\delta g=(g_{1}-g_{2})>0 (<0)$, 

(i) an {\it incommensurate} (c.f the term $\partial_{r}\phi_{\rho}(r)$), gapless
charge-density wave (CDW) results for $\delta g >0$, where the $\phi_{\rho}$ field 
acquires a finite expectation value, and,

(ii) an unconventional SC results for $\delta g <0$, where the {\it dual} field,
$\theta_{\rho}$, becomes non-zero.  

Moreover, the transition between these is of the Kosterlitz Thouless (KT) type.  When 
both couplings are irrelevant, either a singular nFL metal ($\epsilon_{f}=0, V_{fc}=\sqrt{t_{f}t_{p}}$) 
or a 
heavy FL metal ($\epsilon_{f}\ne 0$) results, as derived before.    
 
  Figure (1) ofRef.[8] shows the experimental $T-p$ phase diagram obtained
for $0$GPa $<p< 4.0$GPa~\cite{[8]}. On the basis of our analysis above, we 
suggest that the shape 
of the lines
$T_{vf}(p)$ and $T_{c}(p)$ resemble an underlying KT structure (lines 
in Fig.(1) of Ref.[8]). 
SC clearly involves the $f$-fermion (c.f Eqs.(3)-(9)), and,
with $U_{ff}=\infty$, it is necessarily of the unconventional type.

  The analysis above has interesting consequences, some of which
may be amenable to experimental check for $p>2.0$GPa.  The structure of the low-energy propagators implies~\cite{[19]} that the leading contribution to the optical
conductivity is $\sigma(\omega) \simeq \sigma_{0}/\omega^{1-2\alpha}$, and that
the optical phase angle, $\eta=(1-2\alpha)\pi/2$, near $p_{c2}$.  The corresponding renormalized carrier scattering rate is $1/\tau^{*}(\omega)=-\omega$.cot$(\eta)$.  So the $dc$ resistivity is linear in $T$ as observed near 
$p_{c2}$.  Further, the whole FS is ``hot'', and so disorder will 
not affect the low-$T$ resistivity, as observed~\cite{[8]}. 
 Within DMFT, the electronic Raman intensity will show an anomalous
continuum behavior: $I_{R}(\omega) \simeq \omega^{2\alpha}$.  For $p<p_{c2}$, a low energy pseudogap ($\propto \delta V_{fc}^{2}$) reflecting the UDW metal 
should show up in these responses, while, for $p>p_{c2}$, HFL behavior should
show up below $T_{coh} \propto Z$. 
 Actually, this may be already manifested in extant optical 
measurements~\cite{[vanm],[6]}: it
might be interesting to check whether $\sigma(\omega)$ above the hybridization
gap ($\Delta_{cf}$) indeed follows power-law behavior, as required from our 
analysis.  A slow,
almost power-law fall-off can be readily discerned for the $CeMIn_{5}$ series~\cite{[vanm]}, and such signatures in $CeCu_{2}Si_{2}$ remain to be investigated.
 Magnetic fluctuations should not be affected across $p_{c2}$, as apparently seen~\cite{[8]}.
The abrupt change in the FS volume across $p_{c2}$ should 
manifest in a rapid change in the Hall carrier density, $n_{H}(p,T)$.
The change of sign of $\delta V_{fc}$ across the VF-QPT should drive dramatic 
modification of the Fermi surface across $p_{c2}$: this could be detected by 
dHvA studies. 
  Since, in our approach, incommensurate
 density-wave modes (Luttinger density
bosons) characterize the nFL phase, a study of lattice dynamics across
$p_{c2}$ may uncover their existence: softening of a particular symmetry-adapted phonon mode at an incommensurate wavevector would be clinching evidence.  Finally, if the nFL phase is {\it dual} to the SC in the sense described
above following Eq.(10) (i.e, if one indeed has an underlying KT structure), 
precursor effects above $T_{c}(p)$ in the form of strong vortex liquid 
fluctuations should be visible.  As in the cuprates~\cite{[20]}, Nernst 
experiments might 
be able to confirm this hypothesis.  

  To conclude, a new nFL state, driven by the interplay between non-local
hybridisation ($V_{fc}$) and
local interactions ($U_{fc}$), emerges in the DMFT solution of the EPAM as $V_{fc}$ (pressure)
is tuned across a critical value.  In turn, this nFL state is unstable, either to a HFL, or to
UDW/SC states at low-$T$.  Our mechanism should be relevant to other RE
compounds~\cite{[10],[21]}) 
showing strong VF.

\acknowledgments

  I thank Prof. G. Lonzarich for discussions and his suggestion to look closer at VF in $Ce$-based
systems, and Prof. P. Fulde for a discussion.

\end{document}